\documentclass[conference]{IEEEtran}
\usepackage{cite}
\usepackage{amsmath,amsfonts,amssymb,amsthm, stmaryrd}
\theoremstyle{plain}
\newtheorem{assumption}{Assumption}
\usepackage{algorithmic}
\usepackage{graphicx}
\usepackage{caption}
\usepackage{array}
\usepackage{subfig}
\usepackage{textcomp}
\usepackage{supertabular}
\usepackage{tabularx}
\usepackage{nomencl}
\usepackage{url}
\usepackage{multirow}
\usepackage[draft=False]{hyperref} 
\hypersetup{
     colorlinks=true,
     linkcolor=blue,
     filecolor=blue,
     citecolor=blue,      
     urlcolor=blue,
     }

\usepackage[us]{datetime} 

\newcommand{\dateone}{\formatdate{08}{01}{2018}}
\newcommand{\datetwo}{\formatdate{10}{01}{2018}}

\newcommand{\NRV}{\ensuremath{v}}

\begin{document}

\title{Probabilistic Forecasting of Imbalance Prices in the Belgian Context}

\IEEEoverridecommandlockouts
\IEEEpubid{\makebox[\columnwidth]{978-1-7281-1257-2/19/\$31.00~\copyright2019 IEEE  \hfill} \hspace{\columnsep}\makebox[\columnwidth]{ }}

\author{\IEEEauthorblockN{Jonathan Dumas, Ioannis Boukas, Miguel Manuel de Villena, S\'ebastien Mathieu, Bertrand Corn\'elusse}
\IEEEauthorblockA{Li\`ege University, Montefiore Institute, Belgium\\
		\{jdumas, ioannis.boukas, mvillena, smathieu, bertrand.cornelusse\}@uliege.be}
}

\maketitle

\IEEEpubidadjcol

\begin{abstract}
Forecasting imbalance prices is essential for strategic participation in the short-term energy markets. A novel two-step probabilistic approach is proposed, with a particular focus on the Belgian case. The first step consists of computing the net regulation volume state transition probabilities. It is modeled as a matrix computed using historical data. This matrix is then used to infer the imbalance prices since the net regulation volume can be related to the level of reserves activated and the corresponding marginal prices for each activation level are published by the Belgian Transmission System Operator one day before electricity delivery.
This approach is compared to a deterministic model, a multi-layer perceptron, and a widely used probabilistic technique, Gaussian Processes. 
\end{abstract}


%
\IEEEpeerreviewmaketitle

\section{Introduction}
The progressive large-scale integration of renewable energy sources has altered electricity market behavior and increased the electricity price volatility over the last few years \cite{de2015negative, green2010market, ketterer2014impact}. 
In this context, imbalance price forecasting is an essential tool the strategic participation in short-term energy markets. Several studies take into account the imbalance prices as penalties, for deviation from the bids, to compute the optimal bidding strategy \cite{giannitrapani2016bidding, pinson2007trading, bitar2012bringing, boomsma2014bidding}. However, these penalties are known only a \textit{posteriori}. A forecast indicating the imbalance prices and the system position, short or long, with a confidence interval is a powerful tool for decision making. \textit{Probabilistic} forecasting usually outperforms deterministic models when used with the appropriate bidding strategies \cite{pinson2007trading}.
Whereas the literature on day-ahead electricity forecast models is large, studies about balancing market prices forecast have received less attention. A combination of classical and data mining techniques to forecast the system imbalance volume is given in \cite{garcia2006forecasting}. A statistical description of imbalance prices for shortage and surplus is made by \cite{saint2002wind}. A review and benchmark of time series-based methods for balancing market price forecasting are brought by \cite{klaeboe2015benchmarking}. Both one-hour and one-day-ahead forecasts are considered for state determination, balancing volume, and prices forecasting on the Nord Pool price zone NO2 in Norway.  
%
The contribution of this study can be summarized as follows.
\begin{itemize}
    \item A novel two-step probabilistic approach (TSPA) is proposed for forecasting the Belgium imbalance prices. The TSPA uses a direct forecasting strategy \cite{taieb2012review}. It consists of forecasting an imbalance price for each quarter of the horizon independently from the others, requiring a model per quarter.
    \item It sets a reference for other studies as this subject is rarely addressed.
\end{itemize}

The paper is organized as follows. Section \ref{Section2} formulates the problem. Section \ref{Section3} introduces the novel two-step probabilistic approach and the assumptions made. Section \ref{Section4} describes the numerical tests on the Belgian case. Section \ref{Section5} reports the results. Conclusions are drawn in Section \ref{Section6}. Annex \ref{eem:appendix} provides a short reminder of the imbalance market and the Belgian balancing mechanisms. Notation \ref{Notation} lists the acronyms, parameters, and forecasted or computed variables.

\section{Problem formulation}\label{Section2}

The inputs of the forecasting method are historical and external data, a forecasting horizon $T$, a resolution $\Delta t$, and a forecast frequency. The outputs are the imbalance price forecasts with a confidence interval.  In this study, the input data are the imbalance price history, the NRV, and the marginal prices for activation published by the TSO. The horizon is the time range of the forecasts from a few hours to several hours or days. The resolution is the time discretization of the forecast from a few minutes to several hours. The forecast frequency indicates the periodicity at which the forecasts are computed. For instance, a forecasting module with a six hours horizon, a resolution and periodicity of 15 minutes, computes each quarter, a forecast for the six hours ahead with a 15 minutes resolution.
This paper focuses on the intraday market time scale that requires a forecast horizon from a few minutes to a few hours. The day-ahead time scale requires forecasts of the imbalance prices from 12 to 36 hours, which is not realistic at this stage.

\section{A novel two-step probabilistic approach}\label{Section3}

The probabilistic approach consists of forecasting the imbalance prices in two steps: computing the NRV state transition probabilities, then forecasting the imbalance prices, as depicted in Figure \ref{fig:forecasting_process}. It is motivated by the ELIA imbalance price mechanisms described in Appendix \ref{eem:appendix}.
\begin{figure}[tb]
	\centering
	\includegraphics[width=0.75\linewidth]{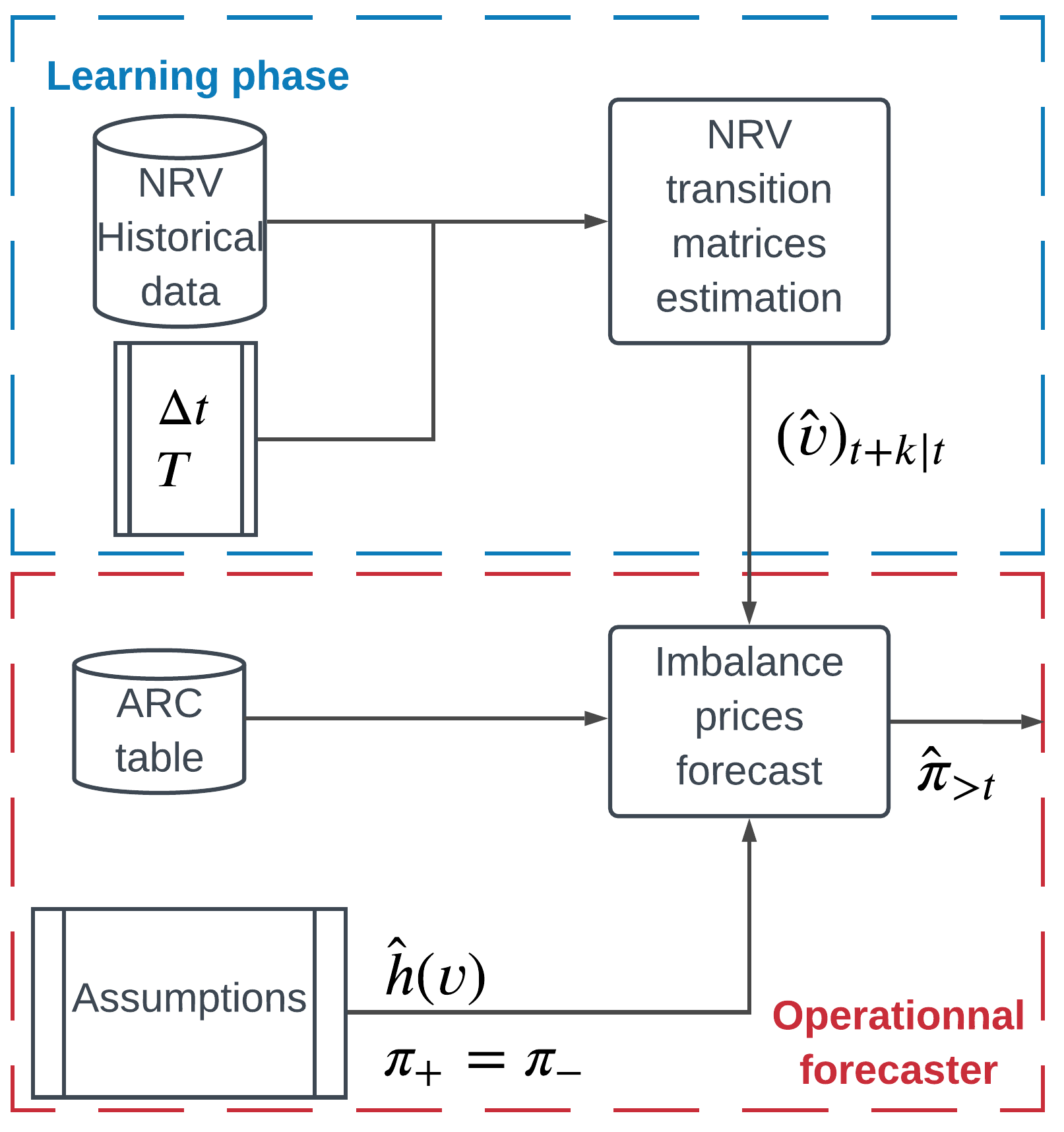}
	\captionsetup{justification=centering}
	\caption{TSPA imbalance price forecasting process.}
	\label{fig:forecasting_process}
\end{figure}

\subsection{Net regulation volume forecasting}

\begin{figure}[tb]
	\centering
	\includegraphics[width=0.8\linewidth]{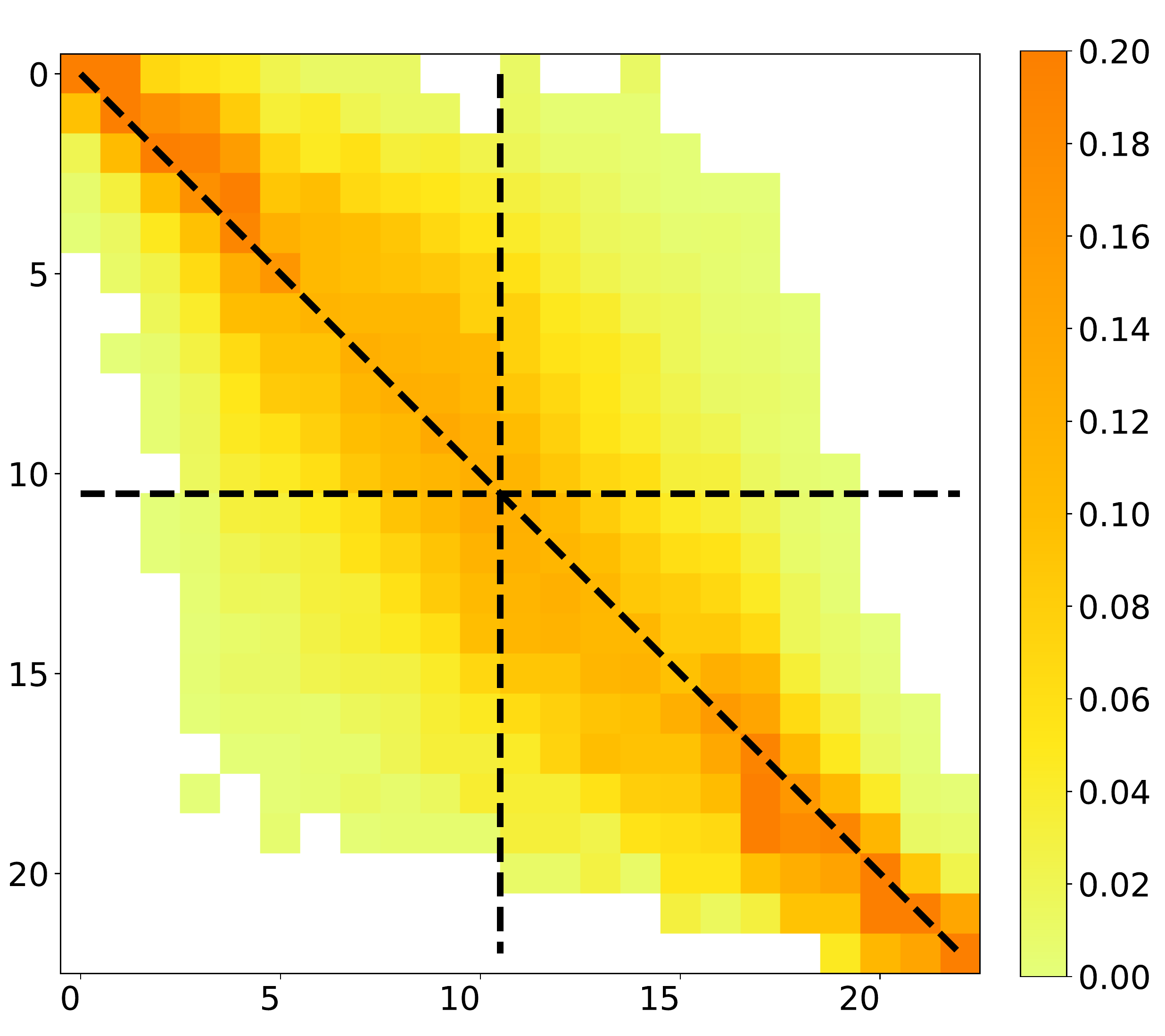}
	\includegraphics[width=0.8\linewidth]{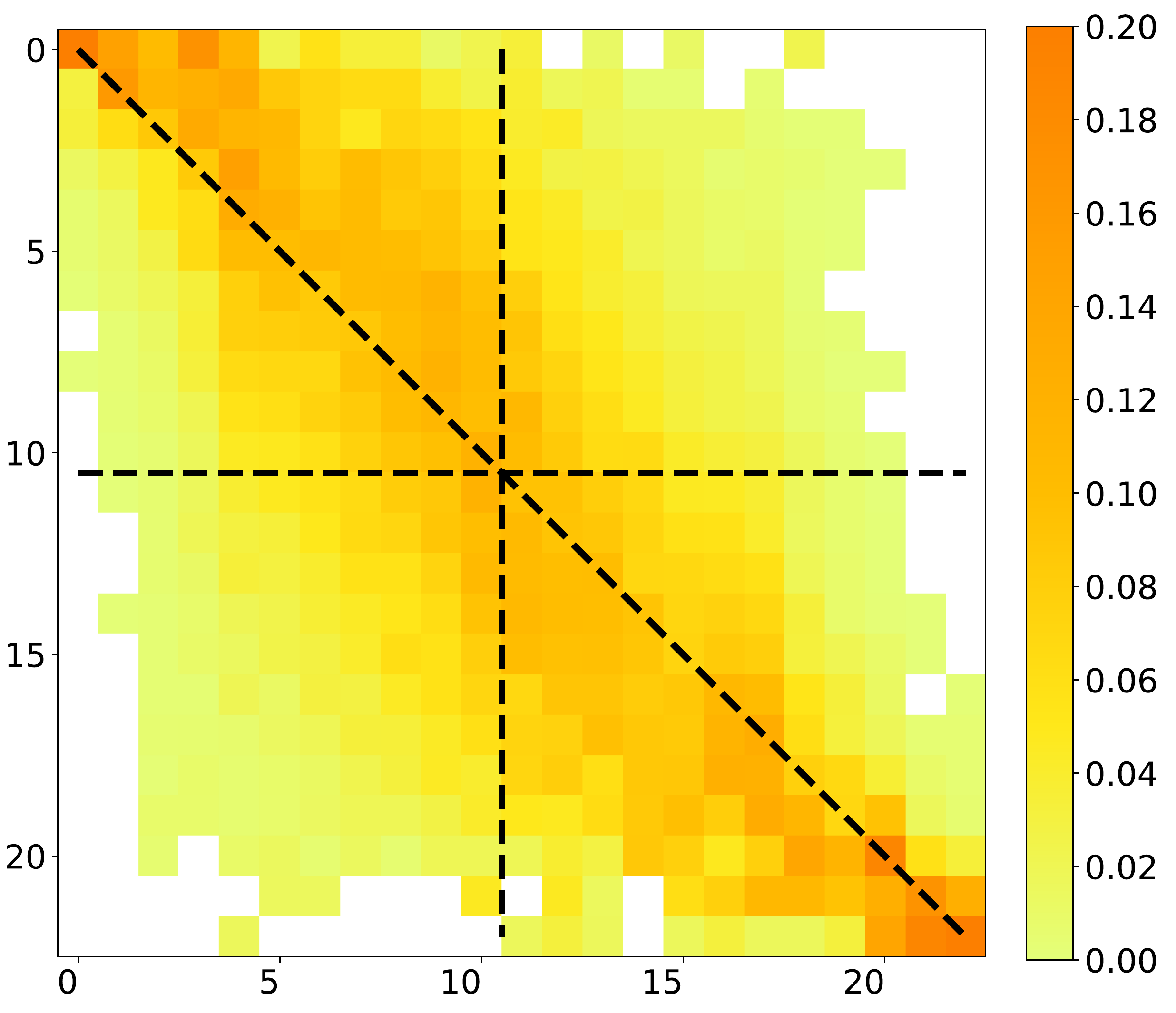}
	\captionsetup{justification=centering}
	\caption{NRV transition matrix from $t$ to $t+15$ min (top) and $t+60$ min (bottom).}
	\label{fig:imbalance-transition_matrix_2017-12_60_heatmap}
\end{figure}

Let consider the $T$ forecasting horizons $k_1=\Delta t, \cdots, k_T = T \Delta t$ with $\Delta t$ the market period, 15 minutes for Belgium. The NRV historical data is discretized into $N$ bins, $\NRV_i$, centered around $\NRV_{i,1/2}$. Note, this discretization has been determined after a statistical study of the NRV distribution.
The $T$ NRV transition matrices $(\NRV)_{t+k|t}$, of dimensions $N \times N$, from a known state at time $t$ to a future state at time $t+k$ are estimated by using the NRV historical data, and referred to as $(\hat{\NRV})_{t+k|t}$. They are composed of the following conditional probabilities $\forall k=k_1, \cdots, k_T$
\begin{equation}
\label{eq:pij_def} 
p^{ij}_{t+k|t}= \Pr[\NRV(t+k) \in \NRV_j \mid \NRV(t) \in \NRV_i ], \  i,j \in {\llbracket 1; N \rrbracket}^2
\end{equation}
with $\NRV(t)$ the measured NRV at time $t$, and $\sum_{j=1}^N p^{ij}_{t+k|t} = 1$ $\forall i \in  \llbracket 1; N \rrbracket$. The conditional probabilities (\ref{eq:pij_def}) are estimated statistically over the learning set (LS) $\forall k=k_1, \cdots, k_T$
\begin{equation}
\label{eq:pij_estimated_def} 
\hat{p}^{ij}_{t+k|t} = \frac{\sum_{t\in LS } 1_{\{ \NRV(t) \in \NRV_i\}}}{\sum_{t\in LS } 1_{\{ \NRV(t+k) \in \NRV_j \ | \ \NRV(t) \in \NRV_i\}}}, \  i,j \in {\llbracket 1; N \rrbracket}^2.
\end{equation}
Figure~\ref{fig:imbalance-transition_matrix_2017-12_60_heatmap} illustrates the matrices $(\hat{\NRV})_{t+k_1|t}$ and $(\hat{\NRV})_{t+k_4|t}$ with 2017 as learning set.
The estimated mean $\hat{\NRV}^m_{t+k|t}$ and standard deviation $\hat{\NRV}^{std}_{t+k|t}$ of the NRV at time $t$ for $t+k$ are calculated as follows 
\begin{equation}
\begin{aligned}
\hat{\NRV}^m_{t+k|t} & = \sum_{j=1}^N \hat{p}^{ij}_{t+k|t}  \NRV_{j,1/2} \\
\hat{\NRV}^{std}_{t+k|t} &= \sqrt{\sum_{j=1}^N \hat{p}^{ij}_{t+k|t} (\NRV_{j,1/2}  - \hat{\NRV}^m_{t+k|t})^2},
\end{aligned}
\end{equation}
with $i$ such as $\NRV(t) \in \NRV_i$.

\subsection{Imbalance price forecasting}

The NRV can be related to the level of reserves activated, and the corresponding marginal prices for each activation level, published by the TSO one day before electricity delivery. We thus first forecast the NRV and its spread among the Gross Upward regulation Volume (GUV) and Gross Downward regulation Volume (GDV). Then we forecast the reserve products activated (contracted or not) to select the most probable MIP and MDP into the ARC table. Finally, the mean and the standard deviation of the imbalance price forecast are derived.

However, the ARC table contains only the contracted reserve products. Most of the time, the first activated reserve products come from the non contracted International Grid Control Cooperation platform (IGCC-/+), the contracted secondary reserve (R2-/+) and the non contracted regulation reserves (Bids-/+)\footnote{Information about the reserve products is available at \url{http://www.elia.be}.}.
For instance, consider a quarter of an hour with an NRV of 150 MW, spread into 170 MW of GUV and 20 MW of GDV. Suppose ELIA activated 80 MW of IGCC+ and 90 MW of R2+. Then, the MIP is given in the marginal activation price of R2+ in the ARC table at the range $[0,100]$ MW. Suppose now that ELIA has activated 20 MW of IGCC+, 20 MW of R2 + and 130 MW of Bids+. Then, the MIP is given in the marginal activation price of Bids+. However, this is not a contracted reserve and its price is not in the ARC table. Then, it is more complicated to predict the MIP and consequently the imbalance prices. Therefore, we introduce several simplifying  assumptions, justified by a statistical study on the 2017 ELIA imbalance data.
\begin{assumption} The NRV is entirely spread into either the GUV (if the NRV is positive) or GDV (if the NRV is negative). \label{assumptionNRV}
\end{assumption} 
The mean and standard deviation of the GUV and GDV are $109 \pm 82$ MW \textit{vs.} $17 \pm 27$ MW when the NRV is positive, while it is $13 \pm 20$ MW \textit{vs.} $110 \pm 73$ MW when the NRV is negative. This assumption enables to select directly in the ARC table the marginal price for activation corresponding to the range of activation equal to the NRV, minus IGCC.
\begin{assumption} The Bids reserve product is not taken into account, thus we suppose that the NRV is spread over the IGCC and reserve products of the ARC table.
\end{assumption}
The percentage of Bids reserve product, positive or negative, activated over each quarter of the 2017 is 11.5 \%.

\begin{assumption} The level of activated IGCC reserve product is modeled by a function $\hat{h}$ of the NRV. 
\end{assumption}
$\hat{h}$ assigns for a given value of NRV a range of activation $p$ into the ARC table. $c_t^p$ is the ARC marginal price at $t$ and for the activation range $p$, with $p \in \llbracket 1; P \rrbracket $. If $\hat{h}(\NRV)$ falls into the activation range $p$, then $c_t^p(\hat{h}(\NRV))$ is equal to $c_t^p$. Due to the 2017 statistical distribution of the IGCC versus the NRV, $\hat{h}$ is defined as follows
\begin{equation}
\hat{h}(x) = \left\{\begin{array}{lcr}
x  & \textnormal{if} & |x| \leq 100, \\
x -100 & \textnormal{if} & x > 100, \\
x + 100 & \textnormal{if} & x < 100.
\end{array} %
\right. 
\label{eq:IGCCmodelisation}
\end{equation}
The mean and standard deviation (MW) of the IGCC+ and IGCC- are
\begin{equation} \notag
 \left\{\begin{array}{lcr}
17 \pm 25 \ \& \ 23 \pm 24  & \textnormal{if} & |NRV| \leq 100, \\
50 \pm 48 \ \& \ 5 \pm 15 & \textnormal{if} & NRV > 100, \\
2 \pm 10 \ \& \ 67 \pm 47   & \textnormal{if} & NRV < 100.
\end{array} %
\right. 
\end{equation}
Generally, ELIA first tries to activate the IGCC product to balance the system. However, when the system imbalance is too high other reserve products are required. 

\begin{assumption} The positive imbalance price is equal to the negative one. 
\end{assumption}
The mean of the positive and negative imbalance prices are $42.23$ and $43.04$ \texteuro$/MWh$. They are different $30.38$ \% of the time, but the NMAE and NRMSE are $0.02$ and $0.06$ \%. Indeed, the positive and negative prices differ only by a small correction parameter if the system imbalance is greater than 140 MW, cf. Appendix~\ref{Belgium Balancing Mechanisms}.

Under these assumptions, the estimated mean $\hat{\pi}^m_{t+k|t}$ and standard deviation $\hat{\pi}^{std}_{t+k|t}$ of the imbalance prices at time $t$ for $t+k$ are calculated as follows
\begin{equation}
\begin{aligned}
\hat{\pi}^m_{t+k|t} & = \sum_{j=1}^N \hat{p}^{ij}_{t+k|t} c_j^{t+k} (\hat{h}(\NRV_{j,1/2})) \\
\hat{\pi}^{std}_{t+k|t} &= \sqrt{\sum_{j=1}^N \hat{p}^{ij}_{t+k|t} (c_j^{t+k}\left(\hat{h}(\NRV_{j,1/2})) -\hat{\pi}^m_{t+k|t} \right) ^2},
\end{aligned}
\end{equation} 
with $i$ such as $\NRV(t) \in \NRV_i$. Finally, on a quarterly basis a forecast is issued at time $t$ and composed of a set of $T$ couples $\hat{\pi}_{>t} := \Big\{ (\hat{\pi}^m_{t+k|t}, \pi^{std}_{t+k|t})  \Big\}_{k=k_1}^{k_T}$.

\section{Tests description}\label{Section4}

This approach is compared to a widely used probabilistic technique, the Gaussian Processes, and a "classic" deterministic technique, a Multi-Layer Perceptron (MLP). Both techniques are implemented using the Scikit-learn Python library \cite{scikit-learn}. The GP uses Mat\'{e}rn, constant and white noise kernels. The MLP has one hidden layer composed of $2 \times n+1$ neurons with $n$ the number of input features. The dataset is composed of the 2017 and 2018 historical Belgium imbalance price and NRV, available on Elia's website. Both the MLP and GP models forecast the imbalance prices based on the previous twenty-four hours of NRV and imbalance prices, representing in total $2 \times 96$ input features. The MLP is implemented with a Multi-Input Multi-Output (MIMO) strategy and the GP with a Direct strategy \cite{taieb2012review}\footnote{Multiple outputs GP regression is still a field of active research \cite{wang2015gaussian,liu2018remarks}.}. The Direct strategy consists of training a model $\hat{f_k}$ per market period $\forall k=k_1, \cdots, k_T$
\begin{equation}
\hat{\pi}_{t+k|t} = \hat{f_k}(\pi_t, \cdots, \pi_{t-k_T}, \NRV_t, \cdots, \NRV_{t-k_T}  ),
\end{equation}
and the forecast is composed of the $T$ predicted values computed by the $T$ models $\hat{f_k}$. In contrast, the MIMO strategy consists of training only one model $\hat{f}$ to directly compute the $T$ values of the variable of interest
\begin{equation}
[\hat{\pi}_{t+k_1|t}, \cdots, \hat{\pi}_{t+k_T|t}]^\intercal = \hat{f} (\pi_t, \cdots, \pi_{t-k_T}, \NRV_t, \cdots, \NRV_{t-k_T}  ).
\end{equation}
For both MIMO and Direct strategies, the forecast is computed quarterly and composed of $T$ values. The forecasting process is implemented by using a \textit{rolling forecast strategy} where the training set is updated every month. The validation set is 2018 where and each month is forecasted by a model trained on a different learning set. For both the MLP and TSPA techniques, the LS size increases by one month each new forecasted month of 2018 with the first LS set to 2017. For the GP technique, the LS is limited to the month preceding the forecast, to maintain a reasonable computation time.

\section{Results}\label{Section5}

The probabilistic forecasts are evaluated using the Pinball Loss Function (PLF) and the Continuous Rank Probability Score (CRPS), and compared to the deterministic ones with the Normalized Mean Absolute Error (NMAE) and the Normalized Root Mean Squared Error (NRMSE) of the mean predicted imbalance prices. The scores $NMAE(k)$,  $NRMSE(k)$,  $PLF(k)$, and $CRPS(k)$ for a lead time $k$ are computed over the entire validation set. The normalizing coefficient for both the NMAE and NRMSE is 55.02 \texteuro $/MWh$, the mean of the absolute value of the imbalance prices over 2018.

\begin{figure}[tb]
	\centering
	\includegraphics[width=3.5in]{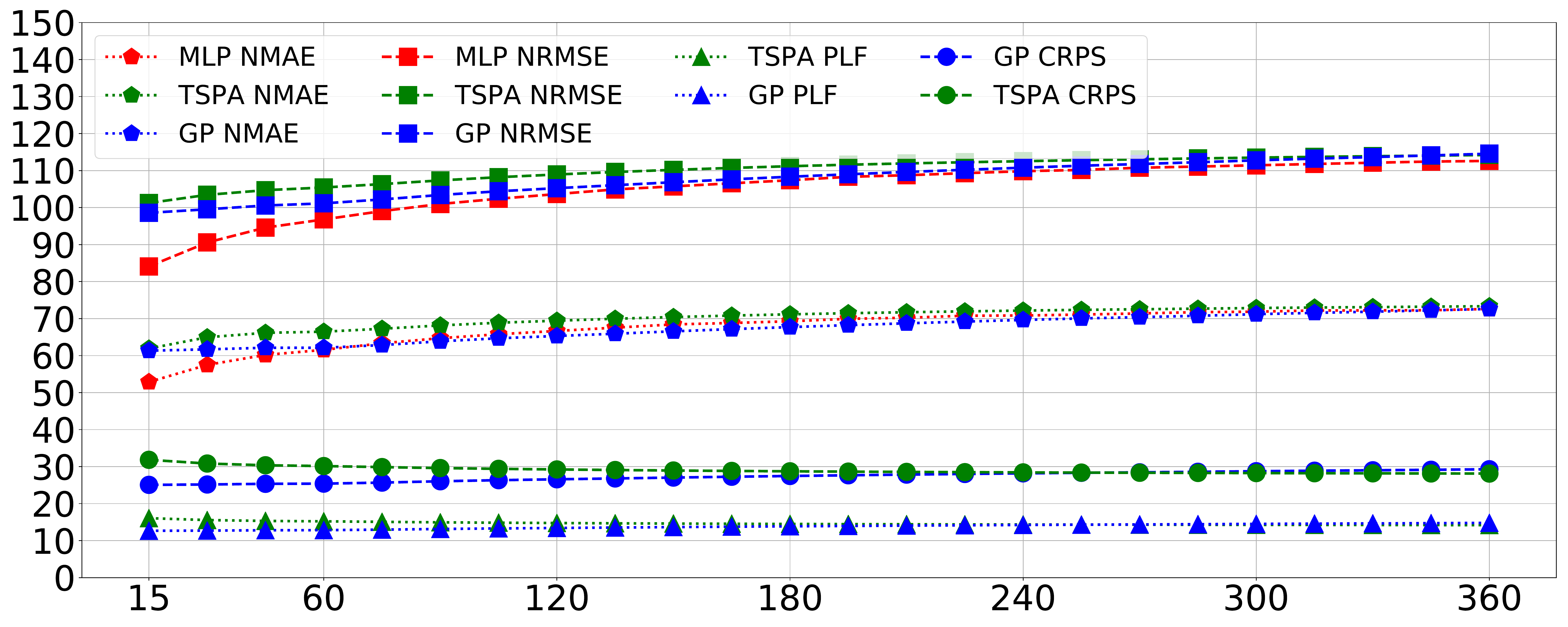}
	\captionsetup{justification=centering}
	\caption{Average scores over all lead times $k$ for each forecasting horizon.}
	\label{fig:scores_per_horizon_M2_GP_MLP_360}
\end{figure}

\begin{figure}[tb]
	\centering
	\includegraphics[width=3.5in]{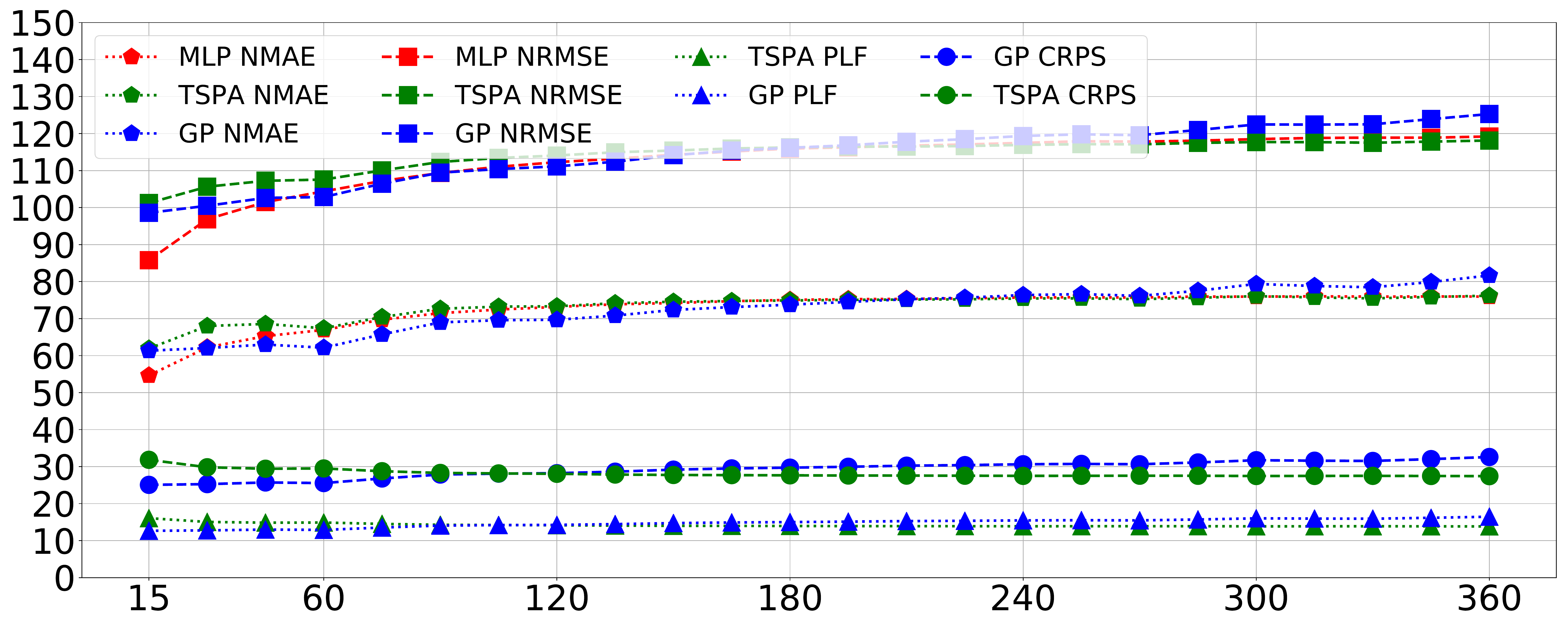}
	\captionsetup{justification=centering}
	\caption{Score per lead time $k$ for the forecasting horizon of 360 minutes.}
	\label{fig:scores_M2_GP_MLP_360}
\end{figure}

\begin{figure}[tb]
	\centering
	\includegraphics[width=1\linewidth]{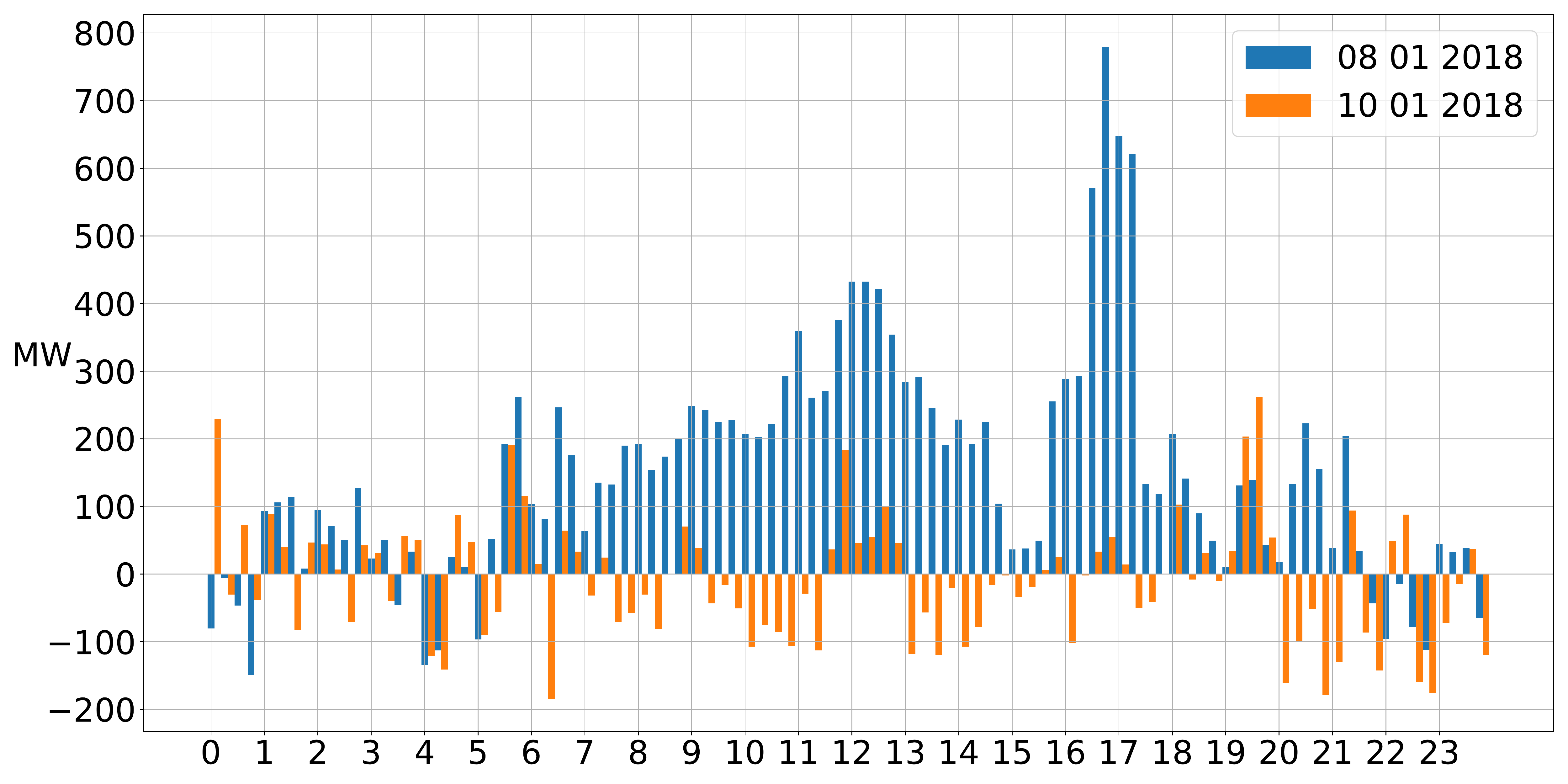}
	\captionsetup{justification=centering}
	\caption{ELIA NRV on $\dateone$ (blue) and $\datetwo$ (orange).}
	\label{fig:nrv08_and_10_012018}
\end{figure}

\begin{figure}[tb]
	\centering
	\includegraphics[width=1\linewidth]{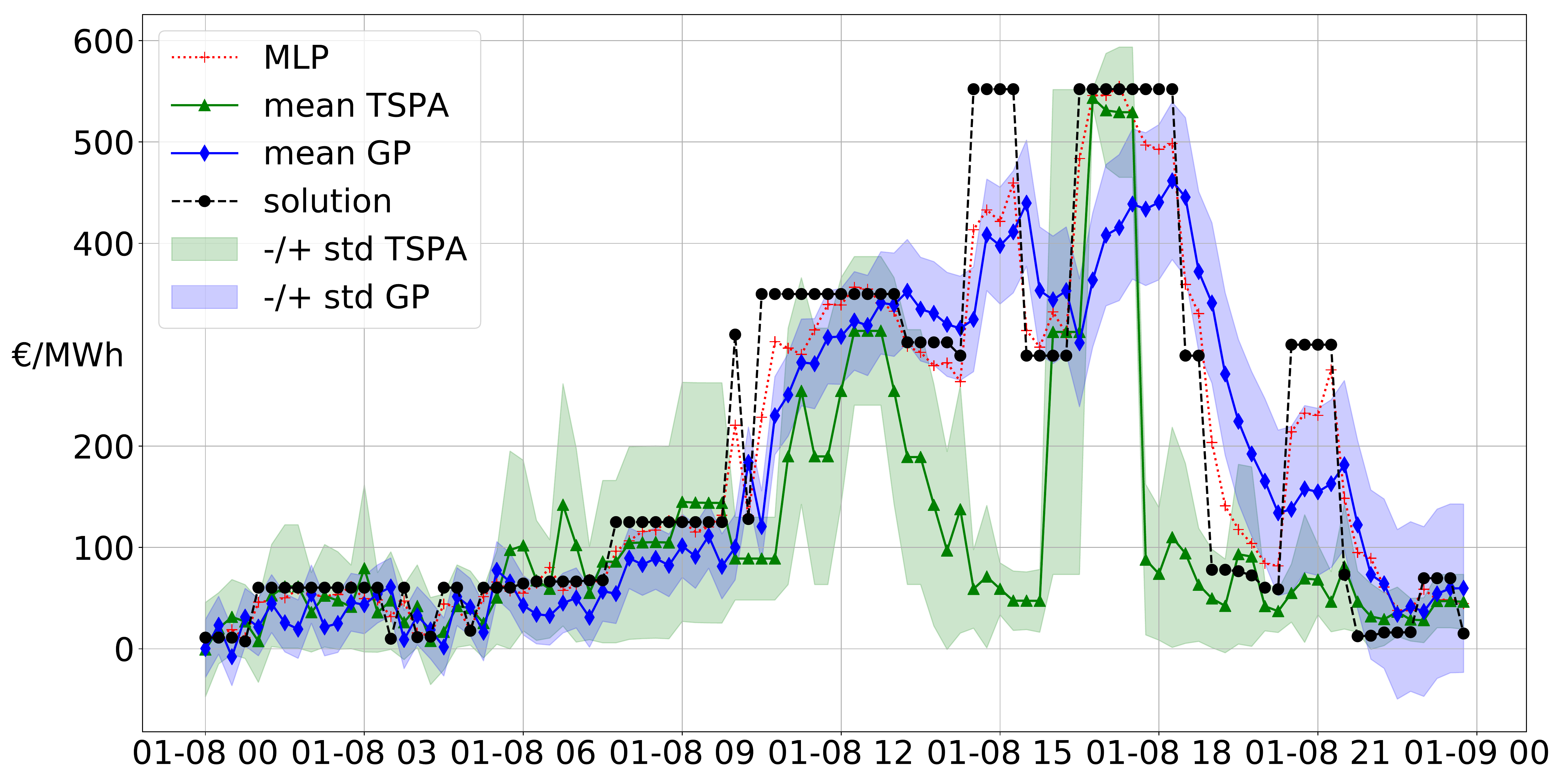}\\[5mm]
	\includegraphics[width=1\linewidth]{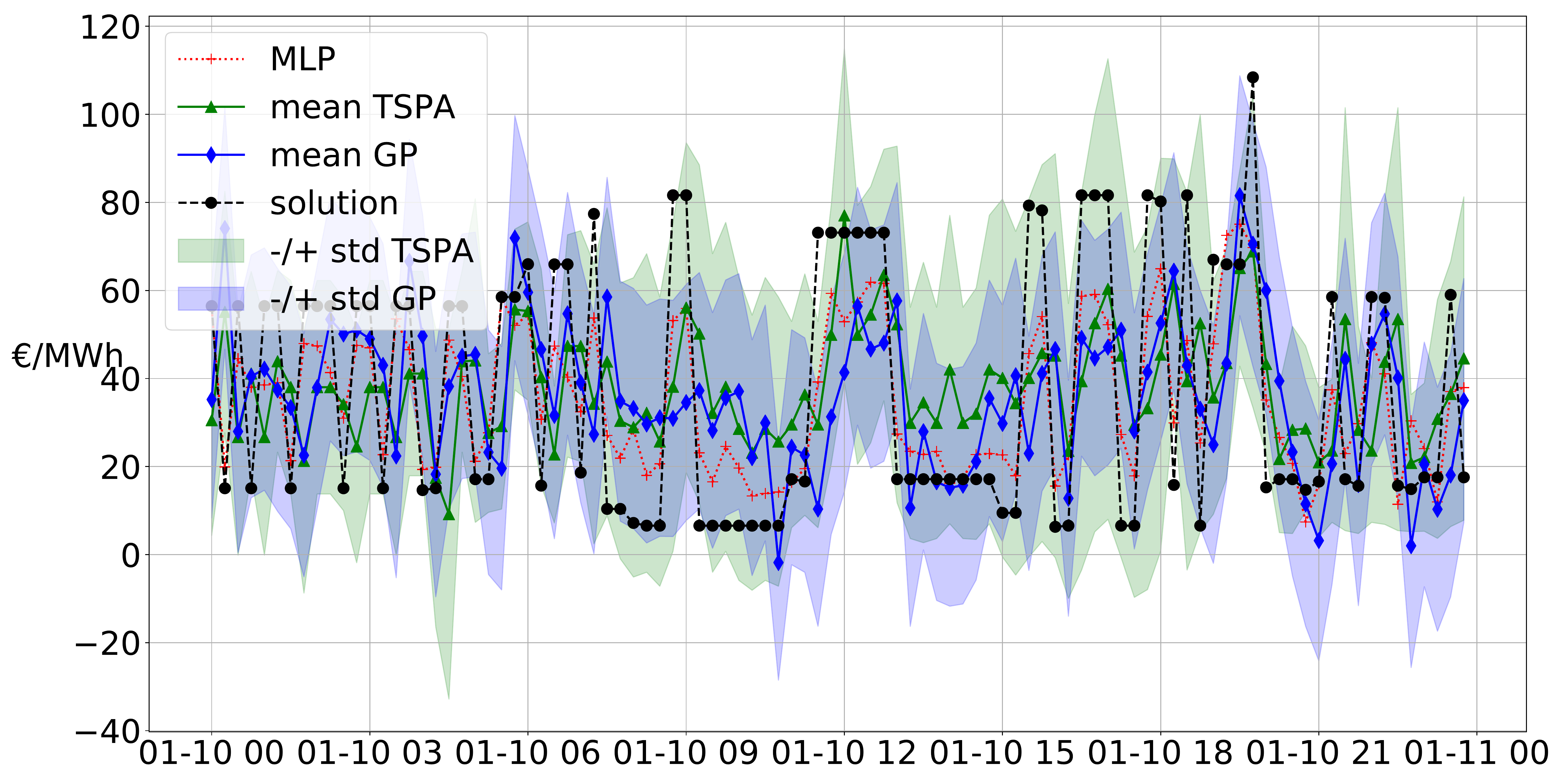}
	\captionsetup{justification=centering}
	\caption{MLP, GP and TSPA 15 minutes horizon forecasts on $\dateone$ (top) and $\datetwo$ (bottom).}
	\label{fig:M2_GP_MLP_15_08012018}
\end{figure}

Table \ref{tab:meanscore} presents the average scores over all lead times $k$ for the horizons of 15, 60 and 360 minutes. Figure \ref{fig:scores_per_horizon_M2_GP_MLP_360} provides the average scores over all lead times $k$ for each forecasting horizon, and Figure \ref{fig:scores_M2_GP_MLP_360} depicts the score per lead time $k$ for the forecasting horizon of 360 minutes.
\begin{table}[tb]
	\begin{center}
	\renewcommand\arraystretch{1.25}
		\begin{tabular}[b]{l l r r r r}
			\hline \hline
			$k$  & Technique & NMAE & NRMSE & PLF & CRPS  \\ \hline
			\multirow{3}{*}{15 min} & MLP  & \textbf{52.74}  & \textbf{84.37}  & -     & -       \\
			 & GP   & 61.33 & 98.59  & 16.48 & 32.64   \\
			 & TSPA & 61.91 & 101.24 & \textbf{16.07} & \textbf{31.84}   \\
			\hline
			\multirow{3}{*}{60 min} & MLP  & \textbf{61.85} & \textbf{97.26}  & -     & -       \\
			 & GP    & 62.13 & 101.14 & 16.09 & 31.87   \\
			 & TSPA  & 66.47 & 105.43 & \textbf{15.22} & \textbf{30.15}   \\
			\hline
			\multirow{3}{*}{360 min} & MLP  & 72.64 & \textbf{112.90} & -     & -       \\
		     & GP   & \textbf{72.61} & 114.56 & 14.79 & 29.29   \\
			 & TSPA & 73.35 & 114.2 & \textbf{14.2} & \textbf{28.12}   \\ \hline \hline
		\end{tabular}
			\caption{Average scores over all lead times $k$.}
		\label{tab:meanscore}
	\end{center}
\end{table}
%
%
%
%
Two days, depicted in Figure \ref{fig:nrv08_and_10_012018}, from the validation set are selected to illustrate the results. On $\dateone$, the ELIA system was short on average, leading to a high NRV and imbalance prices. On $\datetwo$, the ELIA system was alternatively short and long leading to fluctuating NRV and imbalance prices. The 15 minutes horizon forecasts are depicted in Figure~\ref{fig:M2_GP_MLP_15_08012018}, where only the last forecasted value for each quarter is shown. The 60 and 360 minutes horizon forecasts are depicted in Figures~\ref{fig:M2_GP_MLP_60_and_360_08012018} and \ref{fig:M2_GP_MLP_60_and_360_10012018} in Appendix \ref{annex:EEM_figures}. On $\dateone$ the GP provides better results on average as it follows more accurately the actual prices. On $\datetwo$, there is no clear winner. Other figures are reported in Appendix \ref{annex:EEM_figures} for other forecasting horizons.
%
The MLP provides the best NMAE and NRMSE, except for the horizon of 360 minutes, and the TSPA the best CRPS and PLF scores for the three forecasting horizons considered. However, to select the best forecasting model it would be necessary to measure the accuracy of the global bidding chain composed of the forecasting and decision-making modules.

\section{Conclusion and future work}\label{Section6}

This study addressed the problem of forecasting the imbalance prices in a probabilistic framework. The novel two-step probabilistic approach consists of the first step to compute the net regulation volume state transition probabilities. It is used in the second step to infer the imbalance price from the ELIA ARC table and computes a probabilistic forecast. A numerical comparison of this approach to MLP and GP forecasting techniques is performed on the Belgium case. This approach outperforms other approaches on probabilistic error measures but is less accurate at predicting the precise imbalance prices. 

This novel probabilistic approach could be improved by learning models to avoid making our simplifying assumptions, by adding input features to better describe the market situation, and by extending the approach to implement the whole bidding strategy chain, which would allow determining which approach is the best.

\section{Acknowledgments}
The authors thank Haulogy and the Walloon Region for their financial participation.


\bibliographystyle{ieeetr}
\bibliography{biblio}

\section{Appendix}\label{eem:appendix}

\subsection{Balancing mechanisms}\label{Imbalance_mechanism}

A balancing mechanism is designed to maintain the balance over a given geographical area and to control sudden imbalances between injection and off-take. Generally, this mechanism relies on exchanges with neighboring TSOs, the balance responsible parties, and the usage of reserve capacities.
Each party that desires to inject or off-take to the grid must be managed by a Balancing Responsible Party (BRP). The BRP is responsible for balancing all off-takes and injections within its customer's portfolio. The TSO applies an imbalance tariff when it identifies an imbalance between total physical injections, imports, and purchases on the one hand and total off-takes, exports, and sales on the other.
When the BRPs are unable to balance their customer's portfolios, the TSO activates reserves to balance the control area. These reserves are mainly from conventional power plants, which can be quickly activated upward or downward to cover real-time system imbalances. The main types of reserve are the Frequency Containment Reserve (FCR), the Automatic Frequency Restoration Reserve (aFRR), the Manual Frequency Restoration Reserve (mFRR), and the Replacement Reserve (RR). 
The activation of these reserves results from a merit order representing the activation cost of reserve capacity. If the system faces a power shortage, the TSO activates upward reserves that result in a positive marginal price on the reserve market. Then, the TSO pays the Balancing Service Provider. The cost of this activation is transferred to the BRPs. BRPs facing short positions are reinforcing the system imbalance. They must pay the marginal price to the TSO. BRPs facing long positions are restoring the system imbalance. They receive the marginal price from the TSO. This mechanism incentives market players to maintain their portfolios in balance, as well as to reduce the net system imbalance.

\subsection{Belgium balancing mechanisms}\label{Belgium Balancing Mechanisms}
 
This section describes the ELIA imbalance price mechanisms and the data publication that is part of the TSPA inputs. On a 15 minutes basis, the NRV is defined as the sum of the GUV and GDV. The Gross Upward Volume (GUV) is the sum of the volumes of all upward regulations. The Gross Downward Volume (GDV) the sum of the volumes of all downward regulations. If the NRV is positive, the highest price of all upward activated products, the Marginal price for Upward Regulation (MIP), is applied for the imbalance price calculation. If the NRV is negative, the lowest price of all downward activated products, the Marginal price for Downward Regulation (MDP),  is applied. The definition of the positive $\pi_{+}$ and negative $\pi_{-}$ imbalance prices is provided in Table \ref{Elia_imbalance_prices}. The correction parameters $\alpha_1$ and $ \alpha_2$ are zero when the system imbalance is lower than 140 MW and proportional to it when greater than 140 MW.

The MIP and MDP prices are most of the time in the third Available Regulation Capacity (ARC) table. The ARC publication takes into account the applicable merit order, i.e. the order in which Elia must activate the reserve products. Then, within a given priority level, the volumes are ranked by activation price (cheapest first). The marginal price is the highest price for every extra MW upward volume and the lowest price for every extra MW downward volume. The ARC table, showing the activation price of the contracted reserves per activation range of 100 MW, displays the estimated activation price considering a certain NRV. For a given quarter-hour $t$ there are $P$ marginal prices for activation $c^p_t$, $p \in \llbracket 1; P \rrbracket $, each one of them related to the activation range $p$. $P$ is equal to 22 with 11 negatives ranges and 11 positives ranges. The first activation range, $p=1$, corresponds to the interval $[-\infty, -1000]$ MW, the second one to $[-1000, -900]$, ..., $[-100, 0]$, $[0, 100]$ up to $[1000, +\infty]$. The data of day $D$ are published on $D-1$ at 6 pm based on the nomination of day-ahead and intraday programs and bids submitted by the concerned parties. The values, of each quarter hours of the day, are refreshed every quarter-hour. Therefore, the published values are an estimation. However, they are likely to include the MIP and MDP prices at the condition to determine the NRV and its spread between the GUV and GDV. The TSPA takes as input the third ARC table to determine the most probable MIP and MDP prices.
\begin{table}[!htb]
	\begin{center}
		\renewcommand\arraystretch{1.25}
		\begin{tabular}[b]{lll}
			\hline \hline
			BRP perimeter & $NRV < 0$ & $NRV > 0$  \\ \hline
			$ > 0$ & $\pi_{+}  = MDP - \alpha_1 $ & $\pi_{+}  = MIP$  \\ 
			$ < 0$ & $\pi_{-}  = MDP$  & $\pi_{-}  = MIP + \alpha_2$   \\	\hline \hline
		\end{tabular}
		\caption{Elia imbalance prices.}
		\label{Elia_imbalance_prices}
	\end{center}
\end{table}

\subsection{Additional results}\label{annex:EEM_figures}
%
\begin{figure}[!htb]
	\centering
	\includegraphics[width=1\linewidth]{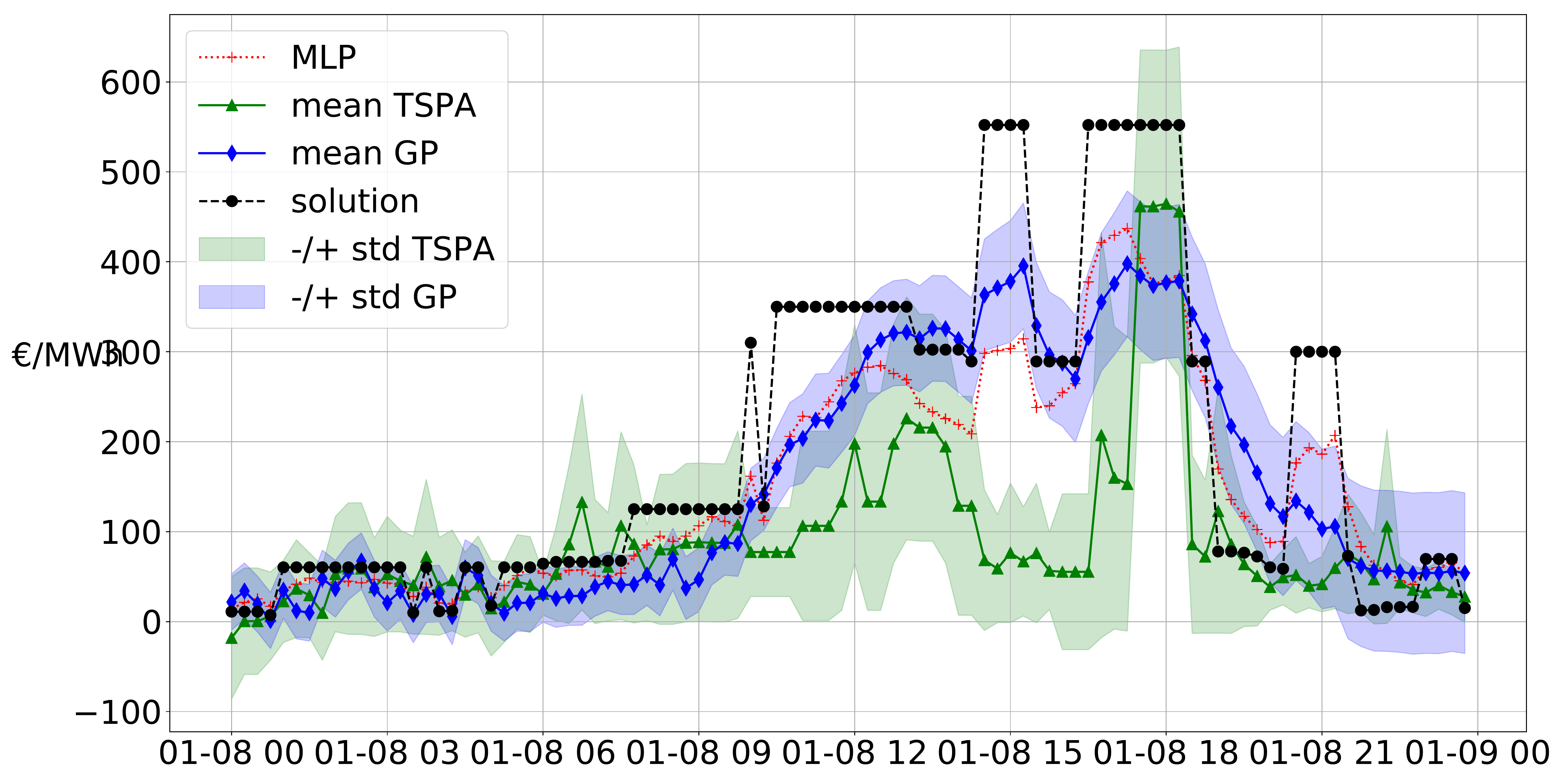}\\[5mm]
	\includegraphics[width=1\linewidth]{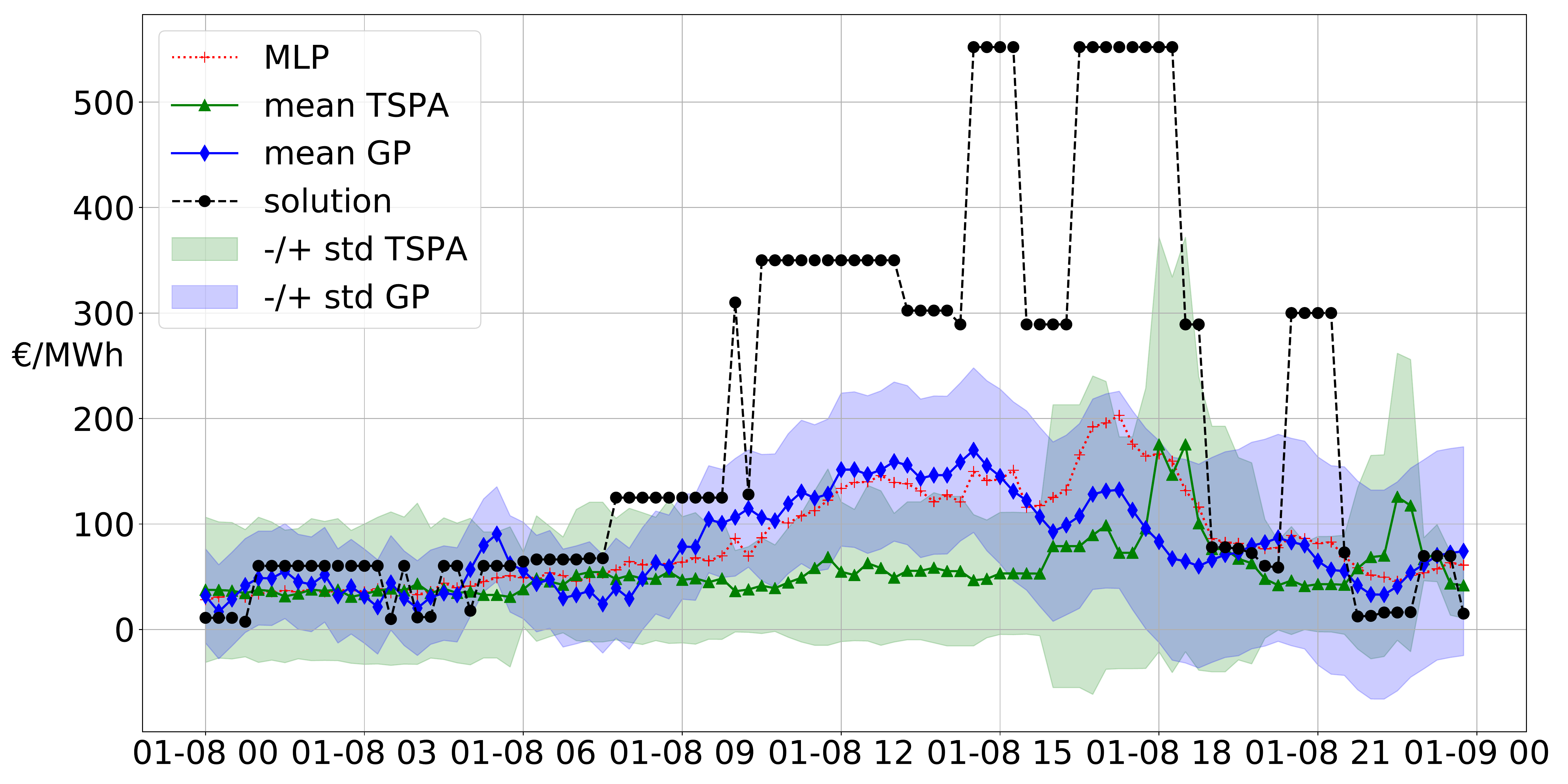}
	\captionsetup{justification=centering}
	\caption{MLP, GP and TSPA 60 (top) and 360 (bottom) minutes horizon forecasts on $\dateone$.}
	\label{fig:M2_GP_MLP_60_and_360_08012018}
\end{figure}
%
%
\begin{figure}[htb]
	\centering
	\includegraphics[width=1\linewidth]{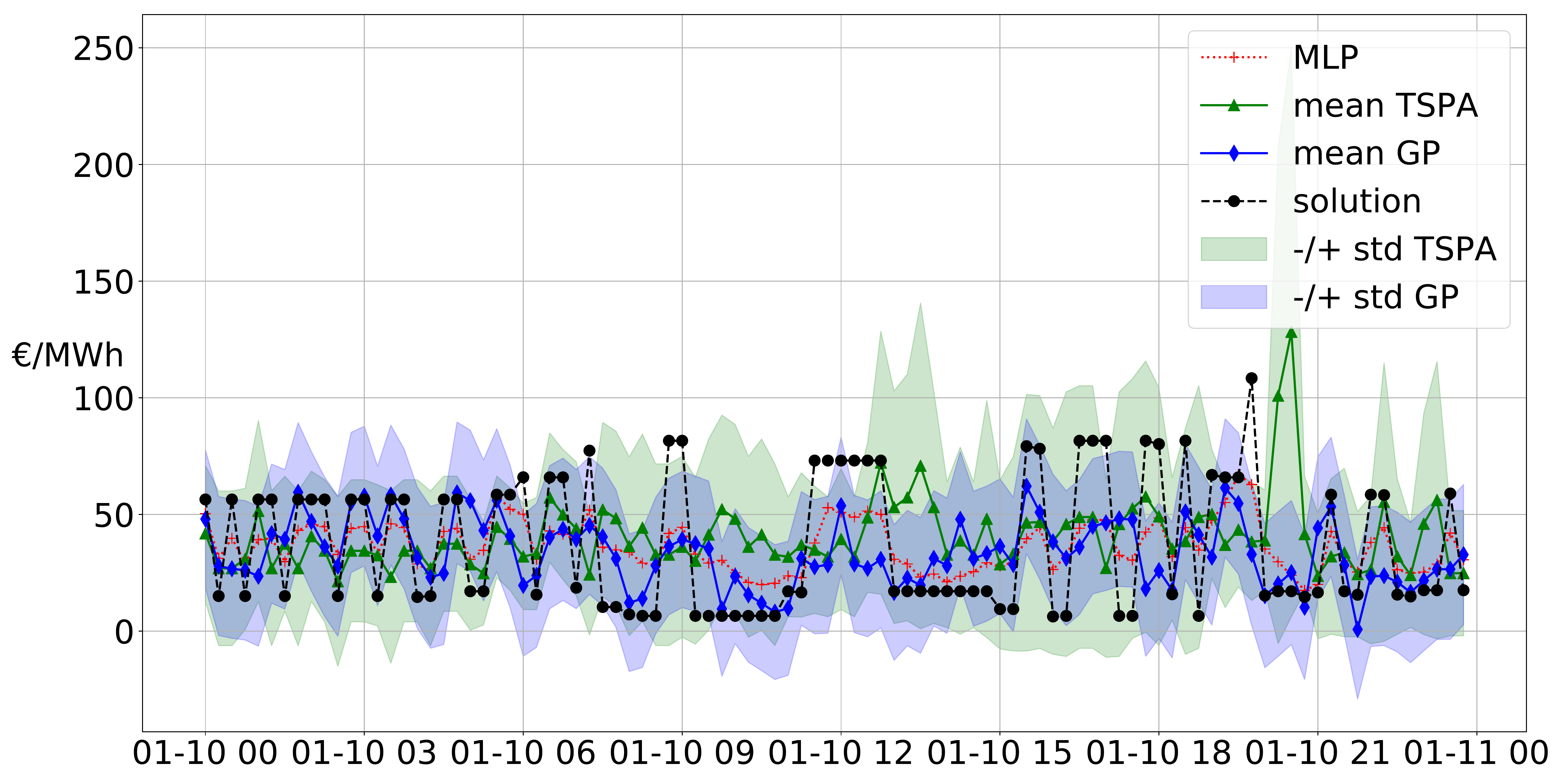} \\[5mm] 
	\includegraphics[width=1\linewidth]{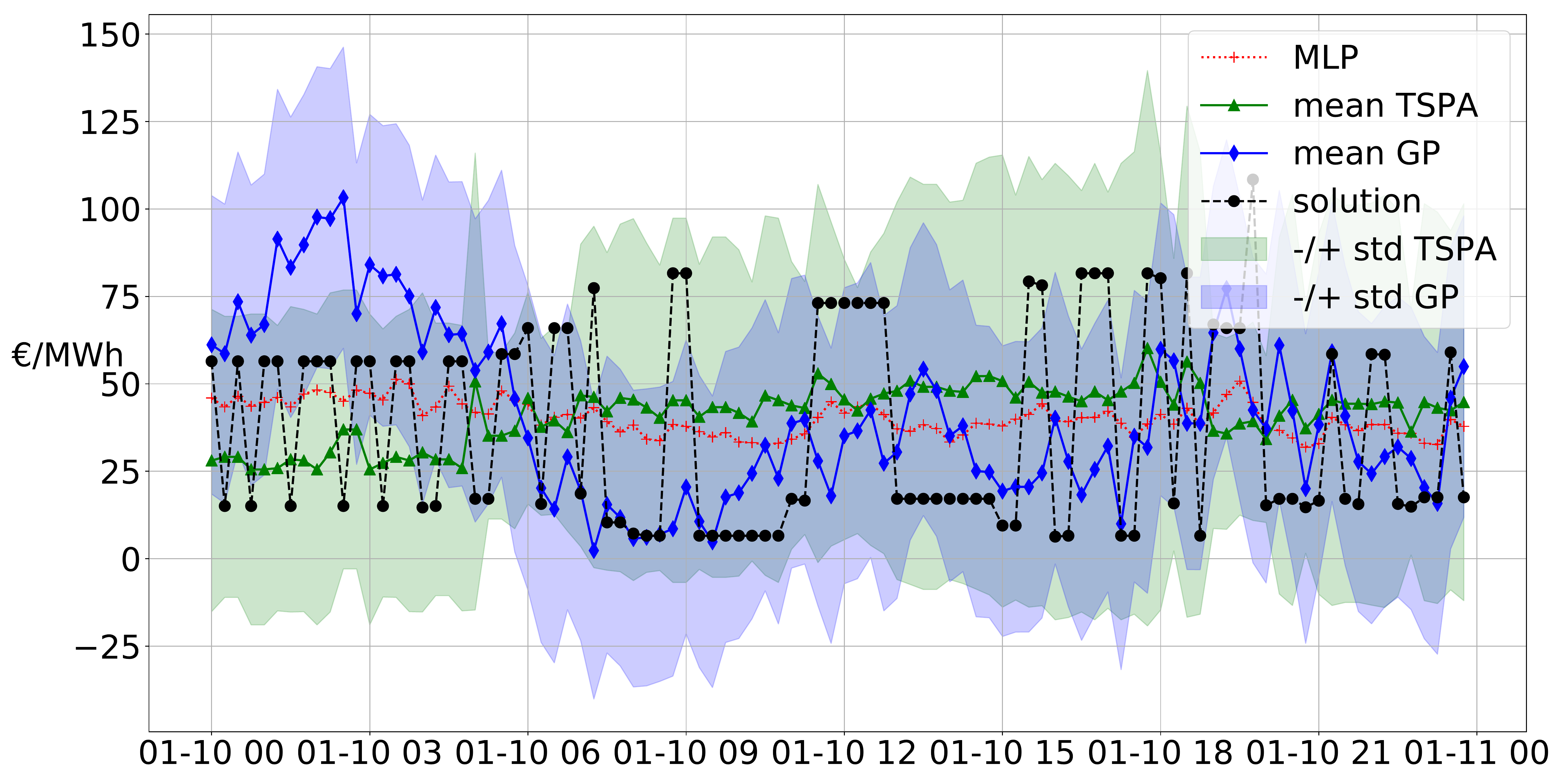}
	\captionsetup{justification=centering}
	\caption{MLP, GP and TSPA 60 (top) and 360 (bottom) minutes horizon forecasts on $\datetwo$.}
	\label{fig:M2_GP_MLP_60_and_360_10012018}
\end{figure}
%
%
\begin{figure}[htb]
	\centering
	\includegraphics[width=1\linewidth]{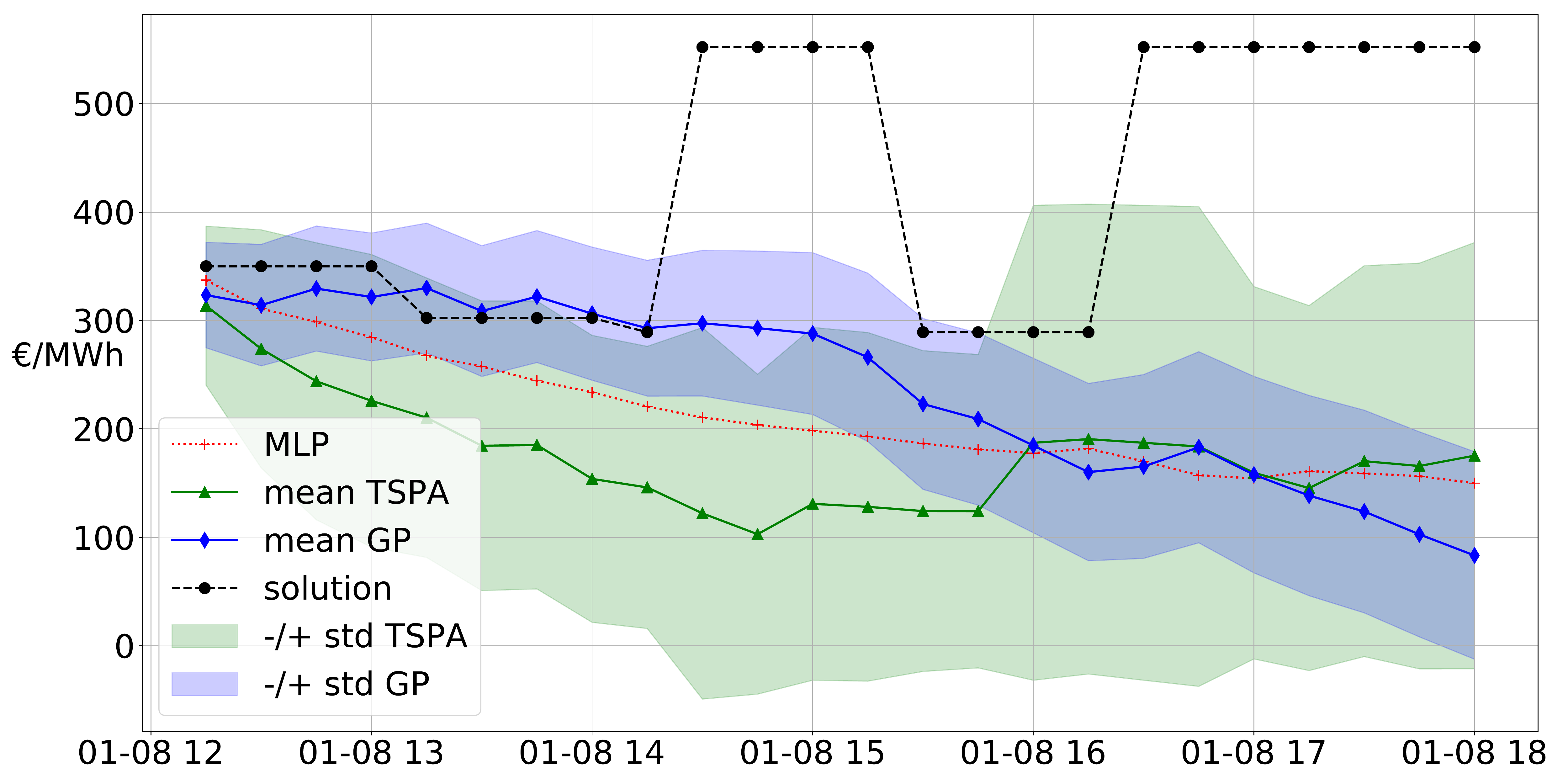}\\[5mm]
	\includegraphics[width=1\linewidth]{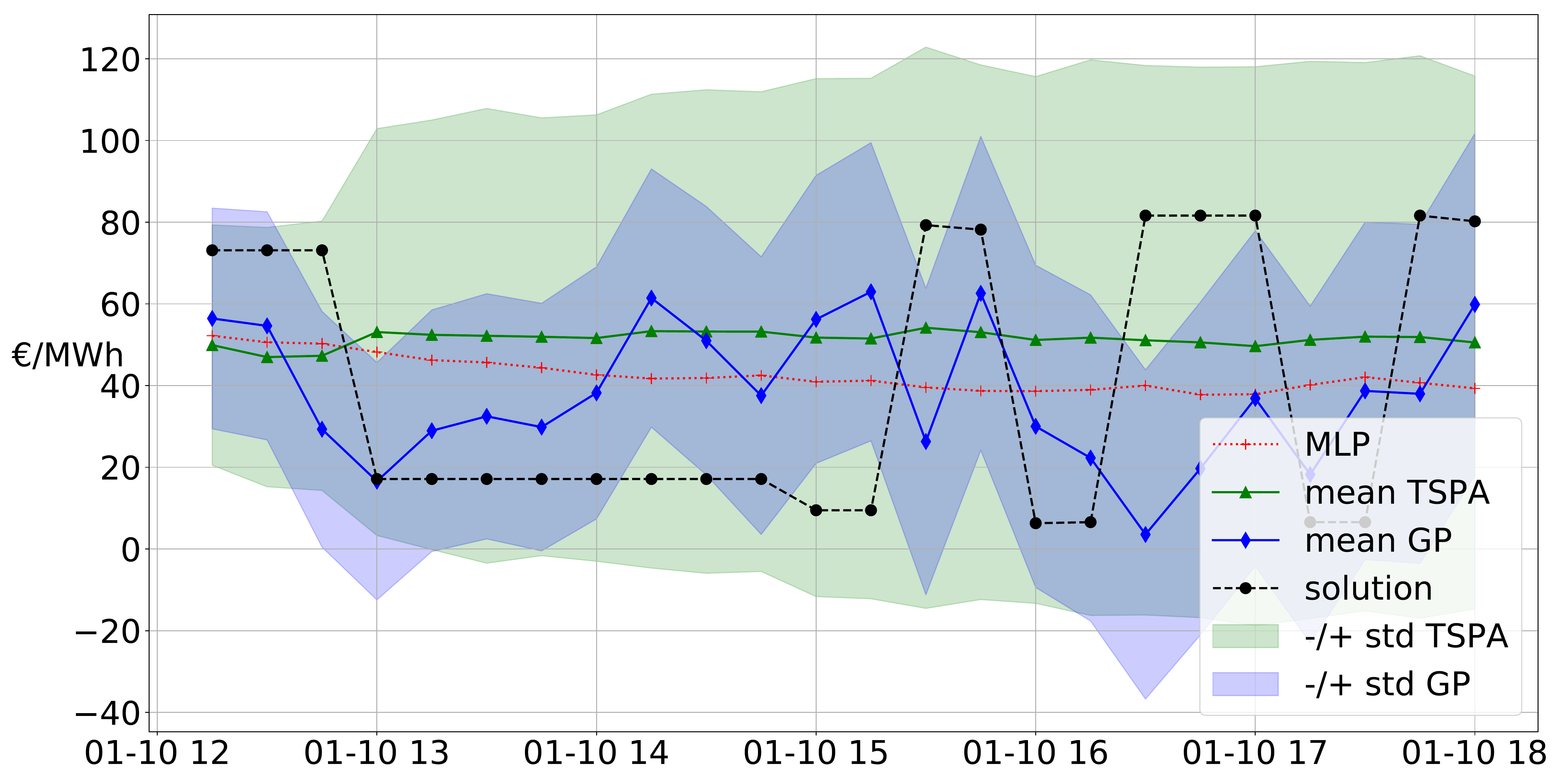}
	\captionsetup{justification=centering}
	\caption{MLP, GP and TSPA forecasts on $\dateone$ (top) and $\datetwo$ (bottom), 12h00 UTC, with an horizon of 360 minutes.}
	\label{fig:M2_GP_MLP_360_08_and_10_012018_1200}
\end{figure}

\clearpage

\section{Notation}\label{Notation}

\subsection*{\textbf{Acronyms}}

\begin{supertabular}{l p{0.75\columnwidth}}
ARC & Available Regulation Capacity \\
BRP & Balancing Responsible Party \\
CRPS & Continuous Rank Probability Score \\
GP & Gaussian Processes \\
GDV & Gross Downward regulation Volume\\
GUV & Gross Upward regulation Volume\\
IGCC & International Grid Control Cooperation \\
MDP & Marginal price for Downward Regulation \\
$\text{metric}$ & NMAE, NRMSE, PLF, CRPS \\
MIMO & Multi-Input Multi-Output \\
MIP & Marginal price for Upward Regulation \\
MLP & Multi-Layer Perceptron \\
NMAE & Normalized Mean Absolute Error \\
NRMSE & Normalized Root Mean Squared Error \\
NRV & Net Regulation Volume \\
PLF & Pinball Loss Function \\
R2 & Secondary reserve, upwards or downwards \\
TSO & Transmission System Operator \\
TSPA & Two-Step Probabilistic Approach \\
\end{supertabular}

\vspace*{1mm}

\subsection*{\textbf{Parameters}}

\begin{supertabular}{l p{0.6\columnwidth} l}
	Symbol & Description & Unit \\
	\hline
	$t$ & Time index  & min\\
	T & Forecasting horizon  & -\\
	$\Delta t$ & Market period  & min\\
\end{supertabular}

\newpage

\begin{supertabular}{l p{0.6\columnwidth} l}
Symbol & Description & Unit \\
\hline
$\pi_{+}$, $\pi_{-}$ & Positive/Negative imbalance price  & \texteuro $/MWh$\\
$ \alpha_1$, $ \alpha_2$ & ELIA parameters for $\pi_{+}$ and $\pi_{-}$ & \texteuro$/MWh$ \\
$c_t^p$ & ARC marginal price at $t$ and for activation range $p$ & \texteuro$/MWh$\\
$\NRV(t)$ & NRV measured at time $t$ & $MW$\\
$\NRV_i$ & NRV bin $i$ & $MW$\\
$\NRV_{i,1/2}$ & Center of NRV bin $i$ & $MW$\\
$(\NRV)_{t+k|t}$ & NRV transition matrix from $t$ to $t+k$ & - \\
$p^{ij}_{t+k|t}$ & NRV conditional probabilities at $t$ for $t+k$  & -\\
\end{supertabular}

\subsection*{\textbf{Forecasted or computed variables}}

\begin{supertabular}{l p{0.6\columnwidth} l}
Symbol & Description & Unit \\
\hline
$\hat{\pi}^m_{t+k|t}$  & Predicted mean imbalance price at $t$ for $t+k$ & \texteuro$/MWh$\\
$\hat{\pi}^{std}_{t+k|t}$ & Standard deviation of $\hat{\pi}^m_{t+k|t}$ at $t$ for $t+k$ & \texteuro$/MWh$\\
$\hat{\pi}_{>t}$ & Set $\Big\{ (\hat{\pi}^m_{t+k|t}, \hat{\pi}^{std}_{t+k|t})  \Big\}_{k=k_1}^{k_T}$ & \texteuro$/MWh$\\
$\hat{\NRV}^m_{t+k|t}$  & Predicted mean NRV at $t$ for $t+k$ & $MW$\\
$\hat{\NRV}^{std}_{t+k|t}$ & Standard deviation of $\hat{\NRV}^m_{t+k|t}$ at $t$ for $t+k$ & $MW$\\
$(\hat{\NRV})_{t+k|t}$ & Estimated NRV transition matrix from $t$ to $t+k$ &- \\
$(\hat{\NRV})_{>t}$ & Set $\Big\{ (\hat{\NRV})_{t+k|t}\Big\}_{k=k_1}^{k_T}$ & -\\
$\hat{p}^{ij}_{t+k|t}$ & Estimated NRV conditional probability at $t$ for $t+k$ & -\\
\end{supertabular}

\end{document}